\documentclass[11pt,twoside]{article}
\usepackage{./asp2014}
\usepackage{comment}

\aspSuppressVolSlug
\resetcounters

\begin{document}

\title{The ngVLA Science Case and Associated Science Requirements}
\author{Eric J. Murphy,$^1$ 
Alberto Bolatto,$^2$ 
Shami Chatterjee,$^3$ 
Caitlin M.~Casey,$^4$
Laura Chomiuk,$^5$
Daniel Dale,$^6$
Imke de~Pater,$^7$ 
Mark Dickinson,$^8$
James Di~Francesco,$^{9,10}$ 
Gregg Hallinan,$^{11}$
Andrea Isella,$^{12}$
Kotaro Kohno,$^{13}$
S.~R.~Kulkarni,$^{11}$
Cornelia Lang,$^{14}$
T.~Joseph~W.~Lazio,$^{15}$
Adam K.~Leroy,$^{16}$
Laurent Loinard,$^{17,18}$
Thomas J.~Maccarone,$^{19}$
Brenda C.~Matthews,$^{9,10}$ 
Rachel A.~Osten,$^{20}$
Mark~J.~Reid,$^{21}$
Dominik Riechers,$^{3}$
Nami Sakai,$^{22}$
Fabian Walter,$^{23}$ \& 
David Wilner$^{21}$}

\paperauthor{Eric J. Murphy}{emurphy@nrao.edu}{0000-0001-7089-7325}{National Radio Astronomy Observatory}{}{Charlottesville}{VA}{22903}{USA}
%\paperauthor{Sample~Author2}{Author2Email@email.edu}{ORCID_Or_Blank}{Author2 Institution}{Author2 Department}{City}{State/Province}{Postal Code}{Country}
%\paperauthor{Sample~Author3}{Author3Email@email.edu}{ORCID_Or_Blank}{Author3 Institution}{Author3 Department}{City}{State/Province}{Postal Code}{Country}

%Please include a brief abstract that will be used by ADS for searching purposes.  
\begin{abstract}
The science case and associated science requirements for a next-generation Very Large Array (ngVLA) are described, highlighting the five key science goals developed out of a community-driven vision of the highest scientific priorities in the next decade.  
Building on the superb cm observing conditions and existing infrastructure of the VLA site in the U.S. Southwest, the ngVLA is envisaged to be an interferometric array with more than 10 times the sensitivity and spatial resolution of the current VLA and ALMA, operating at frequencies spanning $\sim1.2 - 116$\,GHz with extended baselines reaching across North America.  
The ngVLA will be optimized for observations at wavelengths between the exquisite performance of ALMA at submm wavelengths, and the future SKA-1 at decimeter to meter wavelengths, thus lending itself to be highly complementary with these facilities.  
%The result of this exercise is a transformative radio facilities delivering an angular resolution and sensitivity that is each an order of magnitude larger than that of the Very Large Array and the Atacama Large Millimeter Array, allowing it to address fundamental questions in all major areas of astrophysics. 
The ngVLA will be the only facility in the world that can tackle a broad range of outstanding scientific questions in modern astronomy by simultaneously delivering the capability to: (1) unveil the formation of Solar System analogues; (2) probe the initial conditions for planetary systems and life with astrochemistry; (3) characterize the assembly, structure, and evolution of galaxies from the first billion years to the present; (4) use pulsars in the Galactic center as fundamental tests of gravity; and (5) understand the formation and evolution of stellar and supermassive blackholes in the era of multi-messenger astronomy.

\end{abstract}

\section{Introduction}
The Very Large Array (VLA) has had a major impact on nearly every branch of astronomy, and the results of its research are abundant in the pages of scientific journals and textbooks. 
Five years after the completion of the Expanded Very Large Array (EVLA) Project, and more than 40 years since the first VLA antenna was commissioned, the VLA has strengthened its position as the most versatile, widely-used radio telescope in the world. 
Rededicated as the Karl G. Jansky VLA in March 2012, the array continues to make cutting-edge discoveries across a broad range of disciplines including exoplanet formation, galaxy formation in the nearby and distant Universe, and the rapidly growing field of time-domain astronomy. More than 3,000 researchers from around the world have used the VLA to carry out more than 11,000 observing projects.   

Inspired by the VLA's ability to perennially deliver high-impact scientific results, 
%these dramatic results from the VLA, 
and to prepare for the changing landscape in science priorities and capabilities of facilities at other wavelengths, the National Radio Astronomy Observatory (NRAO) recently started to consider ways to continue the legacy of the VLA as one of the most powerful radio telescopes to be included in the next generation of the world's suite of cutting edge astronomical observatories. 
%(e.g., $\sim$30\,m-class optical telescopes, Advanced VIRGO and LIGO, ALMA, LSST, SKA, etc.) 
By teaming with the greater astronomical community in an exercise to develop a cogent science case requiring observations at cm--mm wavelengths, NRAO is now currently pursuing a large collecting area radio interferometer that will open completely new discovery space by delivering an angular resolution and sensitivity that is each an order of magnitude larger than that of the VLA and the Atacama Large Millimeter Array (ALMA), allowing it to address fundamental questions in all major areas of astrophysics.  
%from proto-planetary disks to distant galaxies.  
%The result of this exercise is a transformative radio facilities delivering an angular resolution and sensitivity that is each an order of magnitude larger than that of the VLA and ALMA, allowing it to address fundamental questions in all major areas of astrophysics.  
%Inspired by these dramatic results from the VLA, along with the routine discoveries being uncovered by the Atacama Large Millimeter Array (ALMA), a plan to pursue a large collecting area radio interferometer that will open new discovery space from proto-planetary disks to distant galaxies is being developed by the National Radio Astronomy Observatory (NRAO) and the science community.  

%In this Special Session, invited experts will highlight breakthrough discoveries since the initiation of full science operations with the Karl G. Jansky VLA and will address future transformative scientific opportunities that could be enabled by new capabilities that would maintain the instrument’s pre-eminence well into the 21st century. The importance of combining radio and other wavelength data in the age of multi-wavelength astronomy will be discussed, examining the key contributions of the VLA to the science themes motivating the great observatories that will be commissioned in the next decade. A community-driven vision for the next generation of transformative VLA science will be presented.

This chapter describes the community response following a solicitation from the NRAO to develop key science cases for a future U.S.-led radio telescope, the next generation Very Large Array (ngVLA).  
The resulting list of more than 80 science use cases received by NRAO represent some of the fundamental astrophysical problems that require observing capabilities at cm--mm wavelengths well beyond those of existing, or already planned, telescopes.  
The summary of this exercise has resulted in a transformative radio facility having roughly 10 times the sensitivity of the VLA and ALMA, frequency coverage from $\sim 1.2 - 116$\,GHz with up to 20\,GHz of instantaneous sampled bandwidth, a compact core for good sensitivity to low surface-brightness emission, and extended baselines reaching across North America for extremely high-resolution imaging.  
%of at least hundreds of kilometers and ultimately across the continent to provide high-resolution imaging.  
The ngVLA is being built on the scientific and technical legacies of the VLA and ALMA, and is being designed to provide the next major leap forward in our understanding of planets, galaxies, black holes, and the dynamic sky.
As such, the ngVLA will open a new window on the universe through ultra-sensitive imaging of thermal line and continuum emission down to milliarcecond resolution, as well as deliver unprecedented broad band continuum polarimetric imaging of non-thermal processes.

\section{Developing the ngVLA Key Science Goals}

The ngVLA Science Advisory Committee (SAC)\footnote{\url{http://ngvla.nrao.edu/page/sciencecouncil}}, a group of 24 leading scientists with a wide range of interests and expertise appointed by NRAO, in collaboration with the broader international astronomical community, recently developed a series of than 80 compelling science cases requiring $\approx$200 unique observations between $\sim 1.2 - 116$\,GHz with sensitivity, angular resolution, and mapping capabilities far beyond those provided by the VLA, ALMA, and the Square Kilometre Array Phase 1 (SKA-1). 
The science cases submitted spanned a broad range of topics in the fields of planetary science, Galactic and extragalactic astronomy, as well as fundamental physics, and formed the basis for developing the ngVLA Key Science Goals (KSGs).  
%%%EJM 
%KSGs were identified as projects which satisfied three criteria: (1) the science case addresses an important and currently unanswered question in astrophysics that has broad implications to communities outside of radio astronomy; (2) progress in this area can is uniquely addressed by the capabilities of the ngVLA; and (3) the science case exhibits strong synergies/complementarity with science being pursued by other existing/planned facilities in the >2025 time frame.  

Given the overwhelmingly large spread of compelling science cases generated by the community, it is clear that the primary science requirement for the ngVLA is to be flexible enough to support the wide breadth of scientific investigations that will be proposed by its highly creative user base over the full lifetime of the instrument. 
This mandate is also made obvious given the breadth of scientific endeavors included in this volume, ranging from studies of planet formation and understanding the conditions for habitability in other star systems to rigorous testing of the theory of gravity using pulsars immersed in the space-time potential of the Galaxy's supermassive black hole.   
This in turn makes the ngVLA a different style of instrument than many other facilities on the horizon (e.g., SKA-1, LSST, etc.), which are heavily focused on carrying out large surveys.  

In the next stage of the process, each of the individual science cases were objectively reviewed and thoroughly discussed by the different Science Working Groups within the ngVLA-SAC.
The ultimate goal of this exercise was to distill the top scientific goals for a future radio/mm telescope. 
KSGs were identified as projects which satisfied three criteria: (1) the science case addresses an important and currently unanswered question in astrophysics that has broad implications to communities outside of radio astronomy; (2) progress in this area is uniquely addressed by the capabilities of the ngVLA; and (3) the science case exhibits strong synergies/complementarity with science being pursued by other existing/planned facilities in the $\gtrsim202$5 time frame. 
%The goal of this exercise was to distill the top scientific goals for a future radio/mm telescope. 

The resulting initial KSGs, along with the results from the entire list of over 80 science use cases \citep{usecasesum} were then presented and discussed with the broader community at the ngVLA Science and Technology Workshop June $26 - 29$, 2017 in Socorro NM in an attempt to build consensus around a single vision for the key science missions of the ngVLA.  
Here we describe the five KSGs to come out of this community-driven science use case capture process (see Table~\ref{tab:KSGs}) along with their corresponding requirements that in turn drive the Reference Design described by Selina et al., (this volume, p. \pageref{refdes17}).     

\section{The ngVLA Key Science Goals and Associated Requirements}
In this section we briefly describe each of the five highest-priority ngVLA KSGs that are expected to be carried out during the lifetime of the ngVLA (see Table~\ref{tab:KSGs}).  
We also provide a brief description of their basic requirements that in turn form the foundation used to construct the ngVLA Reference Design described in Selina et al., (this volume, p. \pageref{refdes17}).  
A more detailed description of the ngVLA KSGs can be found in \citep{ngvlaksg17}, and the corresponding full description of the ngVLA Level 0 Science Requirements can be found in \citet{scireq}.  

\begin{table}
    \caption{The ngVLA Key Science Goals}
    \vspace{-6pt}
    \begin{center}
    \begin{tabular}{c|l}
       \tableline
       \noalign{\smallskip}
       {\bf ID} & {\bf Title}\\
       \noalign{\smallskip}
       \tableline
       \tableline
       \noalign{\smallskip}
       KSG\,1 & Unveiling the Formation of Solar System Analogs on Terrestrial Scales\\
       \hline
       KSG\,2 & Probing the Initial Conditions for Planetary Systems and Life with\\ & Astrochemistry\\
       \hline
        KSG\,3 & Charting the Assembly, Structure, and Evolution of Galaxies from the\\ & First Billion Years to the Present\\
       \hline
       KSG\,4 & Using Pulsars in the Galactic Center to Make a Fundamental Test of Gravity\\
       \hline
       KSG\,5 & Understanding the Formation and Evolution of Stellar and Supermassive\\ & Black Holes in the Era of Multi-Messenger Astronomy\\
       \noalign{\smallskip}
       \tableline 
     \end{tabular}
    \label{tab:KSGs}
\end{center}
\end{table}

\begin{figure}[!ht]
\centering
\includegraphics[width=1\textwidth]{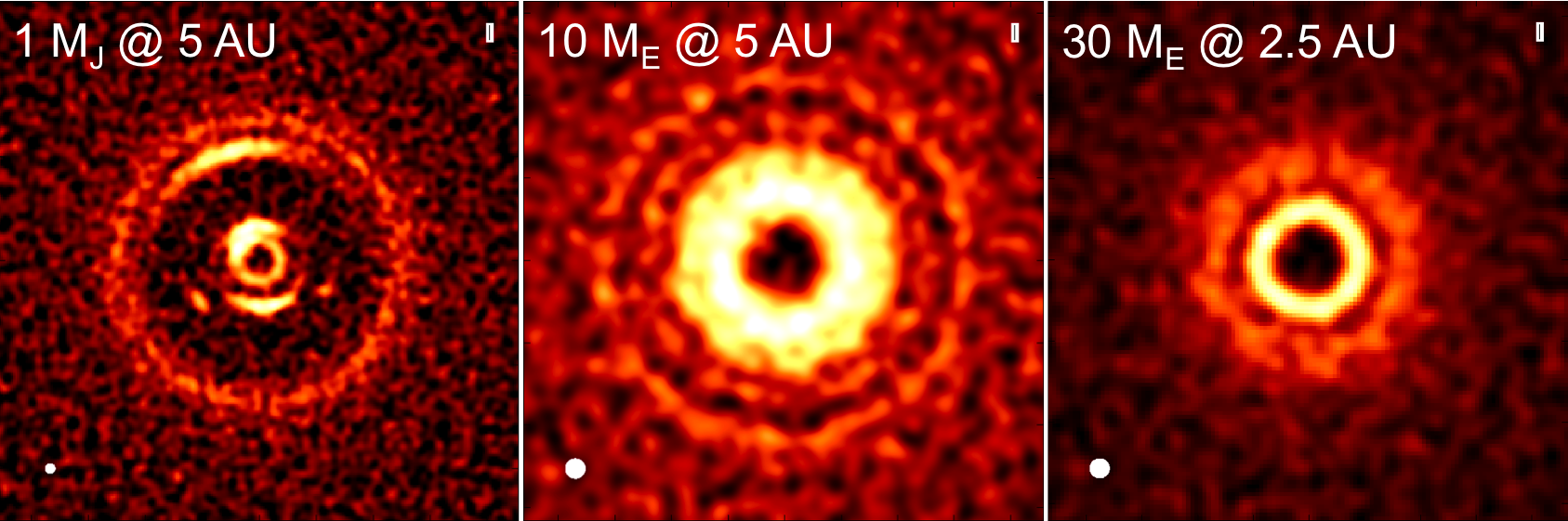}
\caption{\label{fig:ksg1} 
Simulated ngVLA observations of protoplanetary disk continuum emission perturbed by a
Jupiter mass planet at 5\,au (left), a 10 Earth mass planet at 5\,au (center), and a 30 Earth
mass planet at 2.5\,au (right).
The ngVLA observations at 100 GHz were simulated with
5 mas angular resolution and 0.5\,$\mu$Jy/bm rms \citep{lr18}.  }
\end{figure}

\subsection{KSG\,1: Unveiling the Formation of Solar System Analogs on Terrestrial Scales} 
Planets are thought to be assembled in disks around pre-main sequence stars, but the physical processes responsible for their formation are poorly understood. 
Only recently, optical, infrared, and (sub-) millimeter telescopes have achieved the angular resolution required to spatially resolve the innermost regions of nearby protoplanetary disks, unveiling morphological features with characteristic sizes of $>$20\,au suggestive of gravitational perturbations of yet unseen giant planets.  
This in turn provides a powerful tool to measure planet masses, orbital radii, study the circumplanetary environment, and investigate how forming planets interact with the circumstellar material.  
The angular resolution, frequency coverage, and sensitivity of current disk imagery is limited to probing for the presence of planets more massive than Neptune at orbital radii larger than $20 - 30$\,au.  
The next step forward in the study of planet formation is the ability to image the formation of super-Earths and giant planets across the entire disk, particularly within 10\,au from the central star.  
%and to probe for the presence of planets with masses as low as 5 – 10 Earth masses. 

To achieve this science goal requires that the ngVLA have the frequency coverage, sensitivity, and angular resolution to be able to measure the planet initial mass function down to a mass of $5 - 10$ Earth masses.  
This capability will unveil the formation of planetary systems similar to our own Solar System by probing the presence of planets on orbital radii as small as 0.5\,au at the distance of $\approx$140\,pc. 
The ngVLA shall also be able to reveal circumplanetary disks and sub-structures in the distribution of mm-size dust particles created by close-in planets and measure the orbital motion of these features on monthly timescales.  

This in turn requires continuum observations for center frequencies between $20 - 110$\,GHz with angular resolution better than 5\,mas.  
This requirement will enable studies on the formation of planets in the innermost 10\,au of nearby ($\lesssim$140\,pc) proto-planetary disks.  
Extensive simulations of the disks perturbed by planets \citep[see Figure \ref{fig:ksg1};][]{lr18}, suggest that a sensitivity of 0.2\,$\mu$Jy/bm in the continuum at 100\,GHz is required to routinely map structures in the dust distribution created by planets of mass down to 10 Earth-masses and orbital radius of 2.5\,au.
Matching resolution (i.e., 5\,mas) and achieving a continuum sensitivity of order 0.02\,$\mu$Jy/bm at 30\,GHz will map the planet-disk interactions where the disk emission is expected to be optically thin.

\subsection{KSG\,2: Probing the Initial Conditions for Planetary Systems and Life with Astrochemistry}

One of the most challenging aspects in understanding the origin and evolution of planets and planetary systems is tracing the influence of chemistry on the physical evolution of a system from a molecular cloud to a solar system, while also trying to determine the potential for habitability.   
To make significant progress in this area requires that the ngVLA has the frequency coverage and sensitivity to be able to detect predicted, but as yet unobserved, complex prebiotic species that are the basis of our understanding of chemical evolution toward amino acids and other biogenic molecules. 
In doing so, the ngVLA will also allow us to detect and study chiral molecules, testing ideas on the origins of homochirality in biological systems. 
The detection of such complex organic molecules will provide the chemical initial conditions of forming solar systems and individual planets. 

\begin{figure}[!ht]
\centering
\includegraphics[width=1.\textwidth]{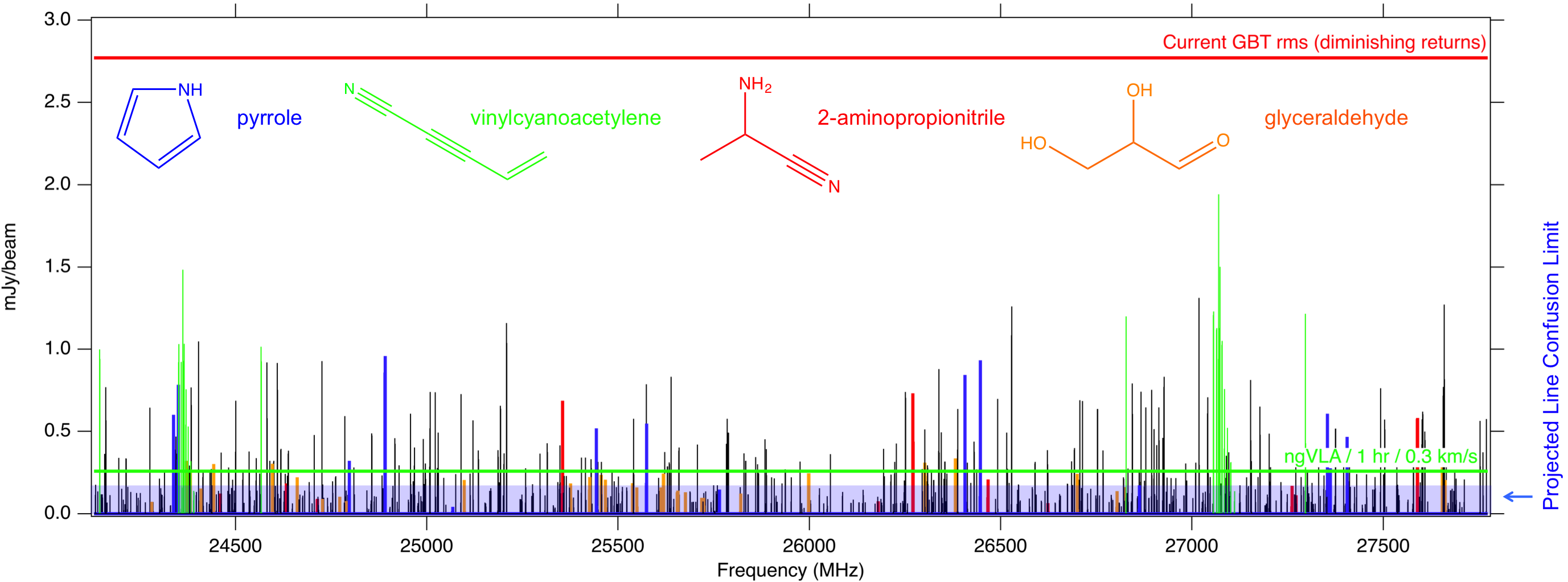}
\caption{\label{fig:ksg2} 
A conservative simulation of 30 as-yet-undetected complex interstellar molecules (black) likely to be observed by the ngVLA above the confusion limit around hot cores with typical sizes of $\sim 1- 4\arcsec$. Key molecules are highlighted in color. 
(Credit: B. McGuire)}
\end{figure}

Presently, existing observations of complex organic (prebiotic) molecules using ALMA and the Great Bank Telescope (GBT) are hitting the limit of what can be accomplished due to a combination of achievable sensitivity  and line confusion at higher frequencies.  
Both problems can be solved by sensitive observations in the cm-wave regime with the ngVLA.  State-of-the-art models predict these molecules will display emission lines with intensities that are easily detectable with the ngVLA, but well below the current detectability thresholds of existing telescopes including ALMA, GBT, and IRAM. 
Figure \ref{fig:ksg2} shows simulations of a representative set of the types of molecules whose discovery will be enabled by the ngVLA: N, O, and S-bearing small aromatic molecules, direct amino acid precursors, biogenic species such as sugars, chiral molecules, and, possibly amino acids themselves.
The simulation assumes column densities of $10^{12} - 10^{14}\,{\rm cm}^{-2}$ (with more complex molecules being assigned lower column densities), a temperature of 200\,K, and 3 km/s linewidth.

To achieve this science goal requires an angular resolution on the order of 50\,mas at 50\,GHz along with an rms sensitivity of 30\,$\mu$Jy/bm/km/s for frequencies between $16 - 50$\,GHz.  
Further, spectral resolution of 0.1\,km/s is required, preferably concurrent with broadband (4+\,GHz) observations.

\begin{figure}[t]
\centering
\includegraphics[width=0.85\textwidth]{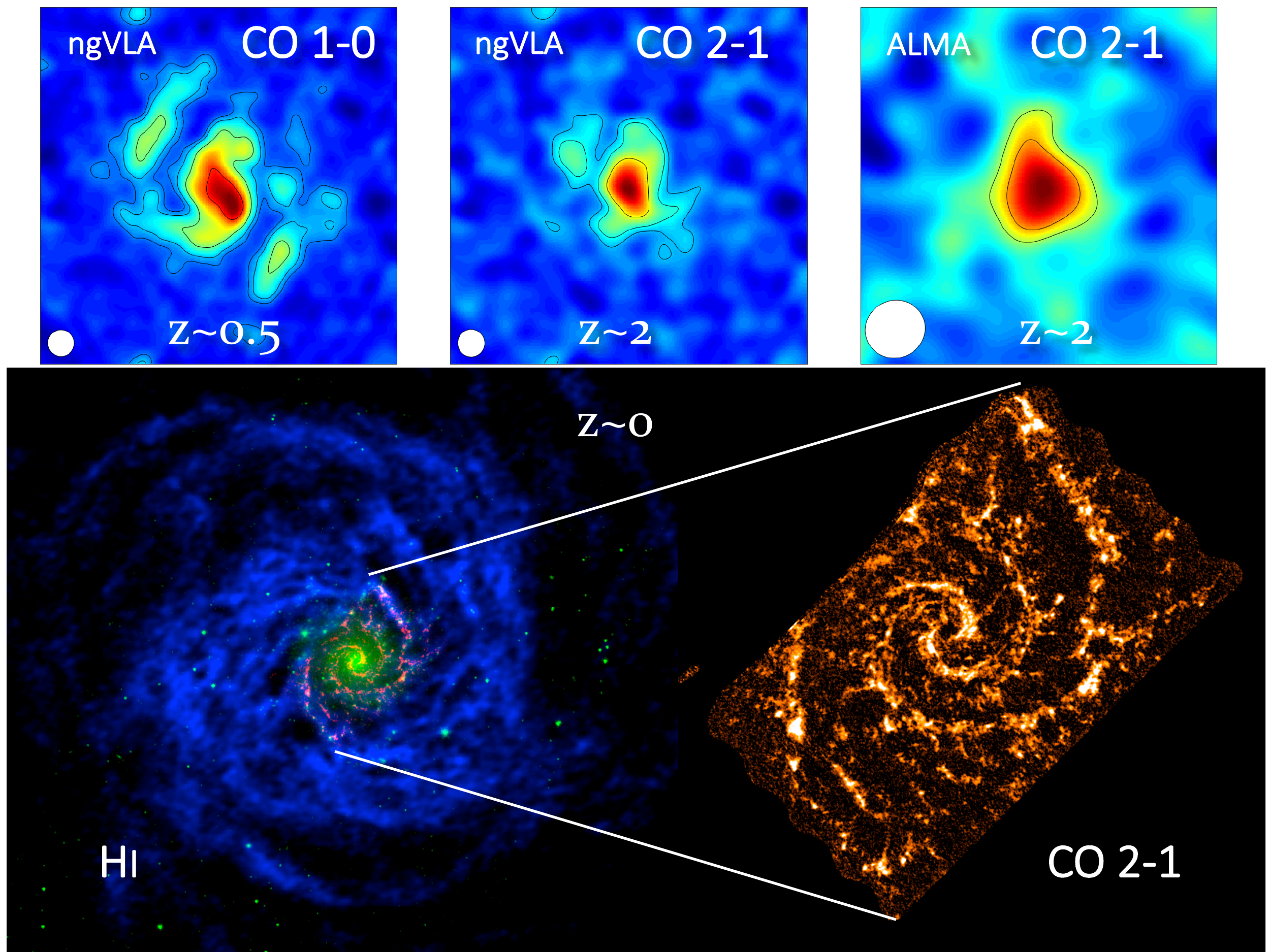}
\caption{\label{fig:ksg3} 
{\it Top Panels:} Simulations based on M\,51 with molecular mass scaled by $1.4\times$ ($z = 0.5$) and $3.5\times$ ($z = 2$) to match the lowest molecular mass galaxies observable by ALMA and NOEMA \citep{memo13}. 
The synthesized beam shown in the bottom left corner is (left to right) $\theta_{s} = 0\farcs19$, 0\farcs20, and 0\farcs43 corresponding to linear scales $L = 1.2, 1.7$, and 3.7\,kpc, respectively.
%The corresponding maximum surface brightness is $T_B  = 6.7, 2.1$, and 1.0\,K, and the black contours enclose regions with $SNR \gtreq 3, 5, 10$, and 20.  
Integration times are 30\,hr.
{\it Bottom Panels:} The spiral galaxy M\,74 illustrating the CO molecular disk imaged by ALMA (red; Schinnerer in prep.), the stellar disk at 4.5\,$\mu$m imaged by {\it Spitzer} \citep[green;][]{sings03}, and the atomic disk imaged in H{\sc i} by the VLA \citep[blue;][]{things08}, showing the gas phases to which the ngVLA will be sensitive. 
{\it Bottom Right Panel:} A zoom in showing the CO $J = 2 \rightarrow 1$ map at 1\arcsec~resolution.
}
\end{figure}

\subsection{KSG\,3: Charting the Assembly, Structure, and Evolution of Galaxies from the First Billion Years to the Present}

To make substantial progress in the field of galaxy formation and evolution requires that the ngVLA have the sensitivity to survey cold gas in thousands of galaxies back to early cosmic epochs, while simultaneously enabling routine sub-kiloparsec scale resolution imaging of their gas reservoirs.  
%This is illustrated by a simulation shown in the top three panels of Figure \ref{fig:ksg3} where the ngVLA 
In doing so, the ngVLA will afford a unique view into how galaxies accrete, process, and expel their gas through detailed imaging of their extended atomic/molecular reservoirs and circumgalactic regions.  
To reveal the detailed physical conditions for galaxy assembly and evolution throughout the history of the universe requires that the ngVLA also have enough sensitivity to map the physical and chemical properties of molecular gas over the entire local galaxy population. 

To carry out detailed studies of CO kinematics of high-$z$ galaxies and blind CO searches of $>1000$ galaxies requires a line sensitivity of $\sim$46\,$\mu$Jy/bm/km/s at 0\farcs1 and 1\arcsec angular resolution between $10 - 50$\,GHz with a spectral resolution of 5\,km/s.   
This is illustrated by a simulation of M\,51 (the Whirlpool galaxy) shown in the top three panels of Figure \ref{fig:ksg3} \citep[see][]{memo13}.  
The spatial and kinematic information recovered by the ngVLA allows for the measurement of a precise rotation curve, which would only be possible to obtain from ALMA with an extraordinarily large ($\sim 1000$\,hr) time investment. 
Furthermore, a large instantaneous bandwidth (i.e., a minimum 1.6:1 BW ratio, up to 20\,GHz instantaneous bandwidth) is required to conduct wide band observations at 5\,km/s resolution to efficiently perform blind surveys of large cosmic volumes in a single observation and provide routine access molecular species in addition to CO (e.g., HCN, HCO$^+$, or N$_{2}$H$^+$). 

Thermal imaging of $0.1 - 0.2$\,K sensitivity of CO (115\,GHz) at 0\farcs1 angular resolution and 1\,km/s spectral resolution is required for detailed studies of molecular gas in the nearby universe (see bottom panels of Figure \ref{fig:ksg3}).  
%A spectral dynamic range of 30 db is also required, while 40 db is desired.  
Thermal imaging of $1 -  5$\,mK sensitivity between 70 and 116\,GHz at $1 - 5\arcsec$ angular resolution and $1 - 5$\,km/s spectral resolution is required to support studies of gas density across the local universe.  
%A spectral dynamic range of 30 db is also required, while 40 db is desired.  

Full $1.2 - 116$\,GHz frequency coverage is required to obtain accurate, simultaneous measurements of star formation rates from free-free continuum and radio recombination line (RRL) emission. 
%A spectral dynamic range of better than 40\,db is required for accurate RRL line-to-continuum ratios. 
Angular resolutions of $0\farcs1 - 1\arcsec$ for continuum imaging at all available frequencies are required. 
A continuum sensitivity of 0.15\,$\mu$Jy/bm at 33\,GHz for a 1\arcsec~synthesized beam is required to robustly study star formation within large samples of nearby galaxies.  
For studies of galaxies in the local universe, accurate recovery of flux density for extended objects on arcminute scales at all frequencies is required, along with the ability to make large mosaics or conduct on-the-fly line and/or continuum mappings of galaxies that extend beyond the area of a single primary beam.  

%Given the expected 33 GHz peak brightnesses within such galaxies, the resulting dynamic range requirement is ~ 37dB.  
Finally, a brightness dynamic range of $\approx$50 and 40\,db is required at 10\,GHz for deep field continuum studies of MW-like galaxies at Cosmic Noon to not be dynamic range limited in total and polarized intensity, respectively.    
Such deep field observations will be sensitive to $\gtrsim$90\% of all stars formed since $z\lesssim3$.

\begin{figure}[t]
\centering
\includegraphics[width=0.7\textwidth]{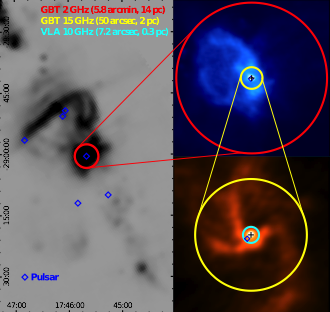}
\caption{\label{fig:ksg4} 
The pulsar distribution near the Galactic Center. Despite being the highest density in the
Galaxy and multiple searches at sensitivities comparable to the VLA, only a few pulsars are
known though $\sim 1000$ are predicted. (Credit: R. Wharton)
}
\end{figure}

\subsection{KSG\,4: Using Pulsars in the Galactic Center to Make a Fundamental Test of Gravity}
Testing theories of gravity requires probing as close as possible to the strong field regime, for which pulsars near the Galactic Center offer a powerful path forward.   
However, only a handful of pulsars in the central half-degree of the Galaxy are currently known (see Figure \ref{fig:ksg4}), which may be the result of 
%likely due to the faintness of pulsars at the distance of the Galactic Center and 
enhanced radio-wave scattering toward the inner Galaxy that further decreases the effective sensitivity of searches, by increasing both the dispersion measure smearing and pulse broadening.  
Observing at higher frequencies than currently planned for the SKA can mitigate radio-wave scattering, but by itself the benefits are limited because of the generally steep radio spectra of pulsars. 
Thus, to achieve this goal, the ngVLA must deliver a combination of sensitivity and frequency range, enabling it to probe much deeper into the likely Galactic Center pulsar population to address fundamental questions in relativity and stellar evolution. 
The ability to address these questions is afforded by the fact that pulsars in the Galactic Center represent clocks moving in the space-time potential of a super-massive black hole and allow for qualitatively new tests of theories of gravity. 
More generally, they offer the opportunity to constrain the history of star formation, stellar dynamics, stellar evolution, and the magneto-ionic medium in the Galactic Center. 
%While there are uncertainties, and the distribution of pulsars could be inhomogeneous, mitigating radio-wave scattering is likely to require a frequency range that includes the lower range anticipated for the ngVLA (> 3 GHz). 

To carryout this science requires that the ngVLA be able to support pulsar search and timing observations from $\sim 1 - 30$\,GHz for Galactic Center pulsars. 
Pulsar searching requires the ability to search on 100\,$\mu$s scales (20\,$\mu$s scales desired), while timing requires 20\,$\mu$s resolution. 
A continuum sensitivity equivalent of order 50\,nJy/bm is desired at 20\,GHz, which is a significant improvement compared to existing 100\,m-class radio telescopes that have found few pulsars, indicating that substantial additional sensitivity is necessary. 
The system timing accuracy also must be better than 10\,ns (1\,ns desired) over periods correctable to a known standard from 30\,min to 10\,yr.  
To efficiently time time multiple pulsars, the array must have the ability to make multiple (minimum 10) beams (i.e., phase centers within the primary beam) within a single subarray, or distributed amongst multiple subarrays.  
Timing multiple pulsars within a single primary beam is desirable. 
Support for 5 or more independent de-dispersion and folding threads is desired.

\subsection{KSG\,5: Understanding the Formation and Evolution of Stellar and Supermassive Black Holes in the Era of Multi-Messenger Astronomy}
While we now know that black holes exist on practically all mass scales, the astrophysics of how these objects form and grow remains a mystery. 
The Laser Interferometer Gravitational-wave Observatory (LIGO) is now detecting black holes that are substantially more massive than previously known stellar mass black holes, and observing black hole-black hole mergers, although we do not know how black hole binaries form. 
While supermassive black holes (SMBHs) are thought to be widespread in galaxy centers, we do not understand how their growth was seeded or how (and how often) these extreme objects merge. 
To address these questions requires that the ngVLA  have the combination of sensitivity and angular resolution to be able to survey everything from the remnants of massive stars to the supermassive black holes that lurk in the centers of galaxies, making it the ultimate black hole hunting machine. 
High-resolution imaging abilities are required to separate low-luminosity black hole systems in our local Universe from background sources, thereby providing critical constraints on the formation and growth of black holes of all sizes and mergers of black hole-black hole binaries. 

To become the ultimate black-hole survey instrument requires that the ngVLA have high angular resolution (mas -- $\mu$as) imaging with relative astrometric accuracy that is $<1$\% of the synthesized beam FWHM or equal to the positional uncertainty in the reference frame, for a bright ($SNR \gtrsim100$) point source.  
%, is required for surveying black holes. 
Such high-resolution imaging will enable proper motion separation of local black holes (both Galactic and in nearby galaxies, out to 15\,Mpc) from background sources. 
Long baselines are required to enable the ngVLA to image the SMBH binaries that will be detected in gravitational waves by LISA and pulsar timing arrays. 
These astrometric science goals benefit from the implementation of very long baselines ($\gtrsim 1000$\,km for mas -- $\mu$as accuracy).  
Associated VLBI recording capabilities shall be available for 3 or more beams (for 2 calibrators and the science target).

\begin{figure}[t]
\centering
\includegraphics[width=0.7\textwidth]{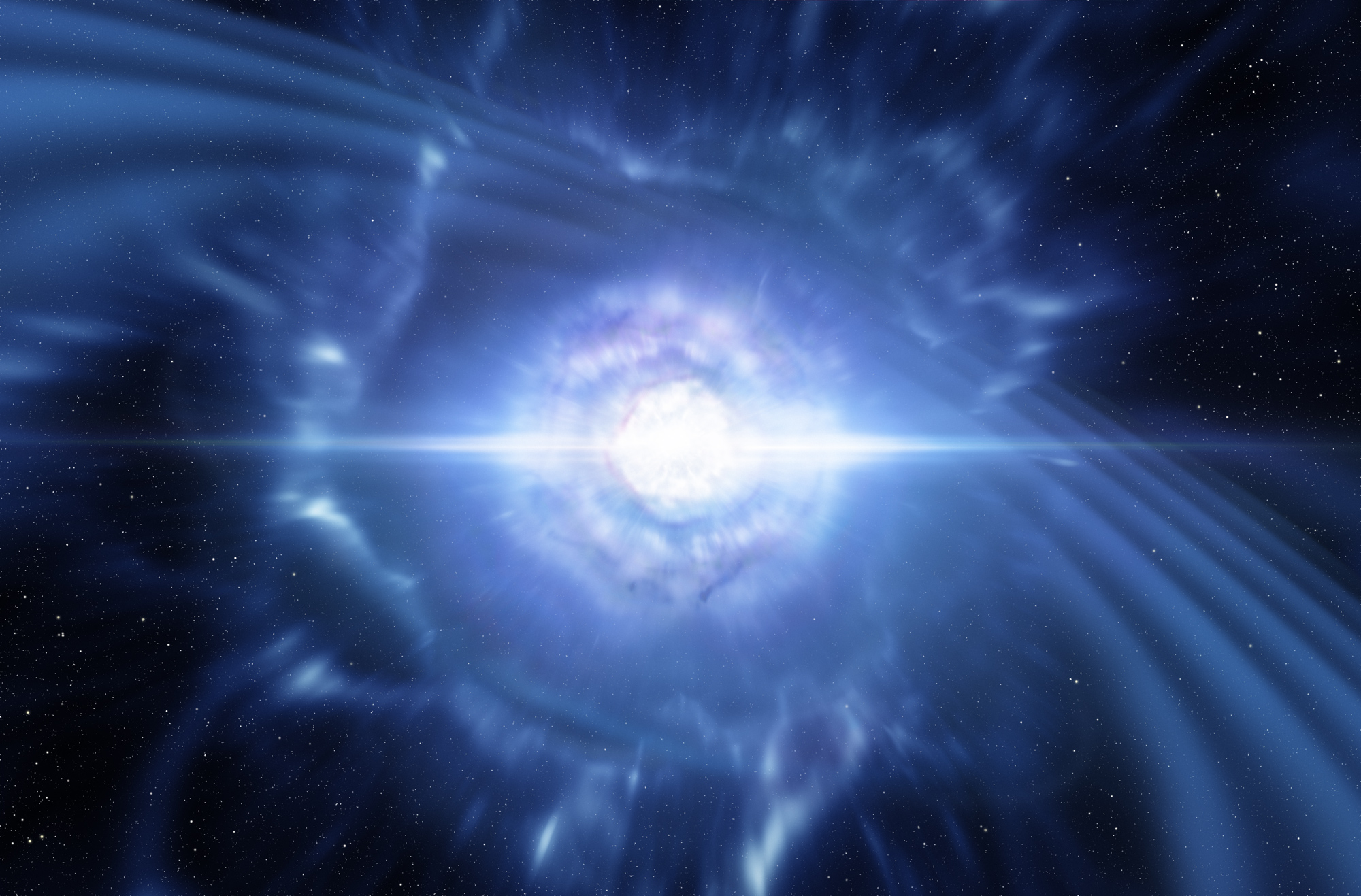}
\caption{\label{fig:ksg5} 
Two tiny, but very dense neutron stars merge and explode as a kilonova. Such a very rare
event produces gravitational waves and electromagnetic radiation, as observed on 17 August
2017. The ngVLA will play a pivotal role in characterizing the physics of such events in the era of multi-messenger astronomy. 
(Artist's impression, Credit: ESO/L. Cal\c{C}ada/M. Kornmesser).  
}
\end{figure}

The field of multi-messenger astronomy continues to mature as we continue to open new astronomical windows through gravitational waves and neutrino observations.  
However, to progress further in our understanding of the physics associated with these phenomena requires the ability to localize and characterize the sources. 
Only the detection of the electromagnetic radiation associated with these energetic, and often cataclysmic events, can provide precise localization, establish energetics and allow us to understand how such events interact with their surrounding environments. 
Thus, to have a transformational impact in the growing era of multi-messenger astronomy, the ngVLA must also be able to identify the radio counterparts to transient sources discovered by gravitational wave, neutrino, and optical observatories (see Figure \ref{fig:ksg5}).   
This requires high-resolution, fast-mapping capabilities to make it the preferred instrument to pinpoint transients associated with violent phenomena such as supermassive black hole mergers and blast waves.

%This is likely to remain challenging, even as other facilities come online over the next decade (Virgo, LISA, KAGRA, LIGO India, PINGU, IceCube Gen2, KM3NeT). Yet it is the combination of multi-messenger information that provides a complete picture of the life-cycle of massive stars, the micro-physics of their explosive deaths, and the formation and evolution of neutron stars, stellar black holes and super-massive black holes. What we know about the electromagnetic counterparts of gravitational waves and neutrino events and their frequency evolution indicates that the ngVLA will be a key component of the rapidly developing landscape of multi-messenger (electromagnetic-neutrino-gravitational wave) astronomy.

Specifically, mapping a $\sim$10 square degree region (i.e., the localization uncertainty expected by gravitational wave detectors when ngVLA is operational) to a depth of $\sim1\,\mu$Jy/bm at 2.5\,GHz for detection of NS-NS and NS-BH mergers is required. 
Completing the on-the-fly mapping of each epoch within $\sim 10$\,hr is desirable. 
Similarly, mapping a $\sim 10$ square degree region at 28\,GHz to a depth of $\sim 10\,\mu$Jy/bm with on-the-fly mapping is required for localization of LISA-detected SMBH mergers.  
Again, completing the on-the-fly mapping of each epoch within $\sim 10$\,hr is desirable. 
Furthermore, an rms sensitivity of $\approx 0.23\,\mu$Jy/bm at 10\,GHz for a 0.6\,mas beam (i.e., on  continental-scale baselines) is required to detect a source like GW170817 with a $SNR \approx 10$ at the Adv. LIGO horizon distance of 200\,Mpc, which will in turn allow for measurements (movies) of its expansion at the 5$\sigma$ level.  

The ability to receive and respond to external triggers rapidly is also an essential requirement to enable multi-messenger science. Triggered response times not to exceed 10 minutes is required, while response times of better than 3 minutes is desired. 
The ability to perform time-domain transient searchers (e.g., for Fast Radio Bursts) requires a search capability on 100\,$\mu$s scales, with 20\,$\mu$s scales desired.

\section{Summary}  
The ngVLA is being designed to tap into the astronomical community's intellectual curiosity by providing them with a world-class instrument that will enable a broad range of scientific discoveries (e.g., planet formation, signatures of pre-biotic molecules, cosmic cycling of cool gas in galaxies, tests of gravity, characterizing the energetics of gravitational wave counterparts, etc.).
Based on community input to date, the ngVLA is the obvious next step to build on the VLA's legacy and continue the U.S.'s place as a world leader in radio astronomy.  
The ultimate goal of the ngVLA is to give the U.S. and international communities a highly capable and flexible instrument to pursue their science in critical, yet complementary ways, with the large range of multi-wavelength facilities that are on a similar horizon.  
Presently, there have been no major technological risks identified.    
However, the project is continually looking to take advantage of major engineering innovations, seeking to optimize the performance and operational efficiency of the facility.  
As the project continues to move forward and mature, the project will continue to work with the community to refine the ngVLA science mission and instrument specifications/performance. 
This science book acts as a major milestones in this effort. 
%This science book and recent reference design study are two major milestones in this effort.  
%through a detailed science book and reference design study.  

\acknowledgements The authors thank the 100's of community members that have contributed to the definition of the ngVLA science mission and corresponding science requirements.  The National Radio Astronomy Observatory is a facility of the National Science Foundation operated under cooperative agreement by Associated Universities, Inc. Part of this work was carried out at the Jet Propulsion Laboratory, California Institute of Technology, under contract with the National Aeronautics and Space Administration.
% ...  % Keep this text on the same line as the \verb"\acknowledgements" command because it makes things a lot easier.

%\bibliography{editor}  % For BibTex

% For non-BibTex:

\clearpage

\affil{$^1$NRAO, Charlottesville, VA, 22903, USA; \email{emurphy@nrao.edu}}
\affil{$^2$Univ. of Maryland, Dept. of Astronomy, College Park, MD 20742, USA}
\affil{$^3$Dept. of Astronomy, Cornell University, Ithaca, NY 14853, USA}
\affil{$^4$Astronomy, UT Austin, 2515 Speedway Blvd Stop C1400, Austin, TX 78712}
\affil{$^5$Dept. of Physics \& Astronomy, MSU, East Lansing, MI 48824, USA}
\affil{$^6$Dept. of Physics \& Astronomy, University of Wyoming, Laramie WY, USA}
\affil{$^7$Dept. of Astronomy and Earth \& Planetary Sciences, UCB, Berkeley, CA, USA}
\affil{$^8$NOAO, 950 N Cherry Avenue, Tucson, AZ 85719, USA}
\affil{$^9$NRC Herzberg, 5071 W. Saanich Rd., Victoria, BC, V9E 2E7, Canada}
\affil{$^{10}$Univ. of Victoria, 3800 Finnerty Road, Victoria, BC, V8P 5C2, Canada}
\affil{$^{11}$Cahill Center for Astronomy \& Astrophysics, Caltech, Pasadena CA, USA}
\affil{$^{12}$Physics \& Astronomy, Rice 6100 Main Street, MS-108, Houston, Texas 77005}
\affil{$^{13}$Inst. of Astronomy, Univ. of Tokyo, Osawa, Mitaka, Tokyo 181-0015, Japan}
\affil{$^{14}$Physics \& Astr., 703 Van Allen Hall, U. Iowa, Iowa City, IA 52242, USA}
\affil{$^{15}$JPL, Caltech, Pasadena, CA 91109, USA}
\affil{$^{16}$Dept. of Astronomy, OSU, 140 West 18th Avenue, Columbus, OH 43210}
\affil{$^{17}$Instituto de Radioastronom\'{i}a y Astrof\'{i}sica, Universidad Nacional Aut\'{o}noma de M\'{e}xico, Morelia 58089, Mexico}
\affil{$^{18}$Univ. Nacional Aut\'{o}noma de M\'{e}xico, Apartado Postal 70-264, 04510 Ciudad de M\'{e}xico, M\'{e}xico}
\affil{$^{19}$Dept. of Physics, Box 41051, Sci. Building, TTU, Lubbock TX, 79409, USA}
\affil{$^{20}$STScI 3700 San Martin Drive, Baltimore, MD 21218, USA}
\affil{$^{21}$ Harvard-Smithsonian CfA 60 Garden Street, Cambridge MA 02138, USA}
\affil{$^{22}$RIKEN, 2-1 Hirosawa, Wako-shi, Saitama 351-0198, Japan}
\affil{$^{23}$MPIfA, K\"{o}nigstuhl 17, Heidelberg D-69117, Germany}

\end{document}